\documentstyle[aps,floats,twocolumn,eqsecnum]{revtex}

\input{psfig.tex}
\begin{document}
\draft
\twocolumn[\hsize\textwidth\columnwidth\hsize\csname @twocolumnfalse\endcsname
\preprint{\vbox{\hbox{CU-TP-864}
                \hbox{CAL-648}
                \hbox{hep-ph/9710337}
}}
\title{Galactic Halo Models and Particle Dark-Matter Detection}
\author{Marc Kamionkowski\footnote{kamion@phys.columbia.edu} and
Ali Kinkhabwala\footnote{ali@rabi.phys.columbia.edu}}
\address{Department of Physics, Columbia University, 538 West
120th Street, New York, New York~~10027}
\date{October 1997}
\maketitle

\begin{abstract}
Rates for detection of weakly-interacting massive-particle
(WIMP) dark matter are usually carried out assuming the Milky Way halo
is an isothermal sphere.  However, it is possible that our halo is
not precisely spherical; it may have some bulk rotation; and the
radial profile may differ from that of an isothermal sphere.  In this
paper, we calculate detection rates in observationally
consistent alternative halo models that
produce the same halo contributions to the local and asymptotic
rotation speeds to investigate the effects of theoretical uncertainty
of the WIMP spatial and velocity distribution.  We use self-consistent
models to take into account the effects of various mass
distributions on the local velocity distribution.  The local halo
density may be increased up to a factor of 2 by flattening or by an
alternative radial profile (which may also decrease the density
slightly).  However, changes in the WIMP velocity distribution in
these models produce only negligible changes in the WIMP detection
rate.  Reasonable bulk rotations lead to only an $O(10\%)$ effect on
event rates.  We also show how the nuclear recoil spectrum in a
direct-detection experiment could provide information on the shape and
rotation of the halo.
\end{abstract}

\pacs{98.70.V, 98.80.C \hfill CU-TP-864, CAL-648, hep-ph/9710337}
]

\def\etal{{et al.}}

\section{INTRODUCTION}

Perhaps the most intriguing explanation for the dark matter in
the Galactic halo is that it is composed of weakly-interacting
massive particles (WIMPs) \cite{jkg}.  These particles typically
have masses between 10 GeV and a few TeV and couple
to ordinary matter only with electroweak-scale interactions.
For example, the leading candidate WIMP is perhaps the
neutralino, the lightest superpartner in supersymmetric
extensions of the standard model \cite{haberkane}.  Several complementary
efforts are currently afoot to detect these halo dark-matter
particles.  For many WIMP candidates, the most promising avenue
is direct detection of the ${\cal O}(10\,{\rm keV})$ recoil energy
deposited in a low-background laboratory detector when a halo
WIMP scatters from a nucleus therein \cite{witten,labdetectors}.
Another promising
technique for many other WIMP candidates is detection of the
energetic neutrinos produced by annihilation of WIMPs which have
been captured in the Sun and/or Earth \cite{SOS}.  There are also
efforts to detect anomalous cosmic-ray positrons,
antiprotons, and gamma rays which may have been
produced by WIMP annihilation in the Galactic halo (see
Ref. \cite{jkg} for a review and further references).

The predicted rates for all of these techniques depend on the
mass and interactions of the WIMP.  The rates for scattering
{}from nuclei also depend on quantities such as quark densities in
the nucleon and on nuclear form factors.
Considerable efforts have been made to
survey the plausible parameter space for supersymmetric WIMPs.
Furthermore, the sources of uncertainty in the predicted
direct-detection and energetic-neutrino rates from, e.g., quark
densities and nuclear form factors have been evaluated and
isolated.

Of course, predictions depend on the spatial and velocity
distribution of WIMPs in the halo.  In most (all?) calculations
of dark-matter detection rates, the halo is assumed to be a
cored isothermal sphere parameterized by a central (or
alternatively, local) density and core radius which are fit to
the observationally inferred halo contribution to the rotation curve.
In this model, the velocity distribution is Maxwell-Boltzmann with a
velocity dispersion determined by the rotation speed at large radii.
Observational uncertainties in the rotation curve and in the
disk and bulge contributions lead roughly to a factor-of-two
uncertainty in the local dark-matter density.  Assuming an
isothermal sphere, one finds the local dark-matter density
$\rho_0=0.2-0.4$~GeV~cm$^{-3}$ and a velocity dispersion $\bar v
= 270 \pm 70$~km~sec$^{-1}$.

In addition to these uncertainties from the rotation curve and
disk mass distribution, deviations
{}from the standard nonrotating isothermal spherical halo are also
plausible, if not probable.  Essentially all the empirical
information we have on the halo is provided by the rotation
curve.  To a first approximation, almost any halo mass
distribution which gives rise to a flat rotation curve is
acceptable.  Although there are some arguments that
the halo must be more diffuse than the disk \cite{kulkarni}, there is
no reason why it
should be perfectly spherical.  In fact, there is ample
evidence that the halos of several external spiral galaxies are
flattened by roughly a factor of two \cite{rix} and now some
evidence that the Milky Way halo is similarly flattened
\cite{rob}.  The dominant
effect of flattening on the detection rate is through the local
dark-matter density \cite{gyuk}.  However, flattening may also
affect detection rates through the velocity distribution,
which has not been taken into account.

Bulk rotation can also affect the
velocity distribution of WIMPs seen at the Earth.
Again, the rotation curve is determined by the halo mass
distribution and is insensitive to its velocity distribution.
Therefore, there is no empirical evidence to rule out a halo
with some bulk rotation.  Although there are theoretical
arguments against a rotation-dominated velocity distribution,
there are also reasons to expect the halo to have some bulk rotation
\cite{frank,barnes}. 

There may also be theoretical uncertainties in the halo radial
profile.  The functional form for the radial profile commonly
assumed is in fact a phenomenological model which produces a
linearly rising rotation curve at small radii and a flat
rotation curve at large radii.  There are other radial profiles
which will satisfy these requirements and produce the same
rotation speeds at the Galactocentric and large radii to which
the models are fit.

In this paper, we investigate uncertainties in the WIMP detection rate
which arise from imprecise knowledge of the spatial and velocity
distribution of dark-matter particles.  To do so, we use a class of
self-consistent models for a flattened and/or rotating halo which have
been developed by Evans \cite{evans}, and consider several plausible
spherical distributions.  All the models we consider produce the same
halo contribution to the local and asymptotic rotation speeds.
Some of our models may appear to be extreme (in
terms of flattening, bulk rotation, or central density) to some
Galactic-dynamics experts; however, our primary aim is to
provide a {\it conservative} estimate of the uncertainty in
dark-matter detection rates from  uncertainties in the halo
distribution, and the models we consider span a range of
observationally plausible---though not necessarily theoretically
favored---models.

We find that flattening and/or changes to the radial profile may
increase the density by roughly a factor of two.  However,
either departure {}from the canonical isothermal sphere has a
negligible effect on the  velocity-distribution dependence of
the event rate.  The bulk rotations which may arise in realistic
galaxy-formation scenarios will have no more than a 10\% effect
on detection rates.

In the next Section, we review the procedure for calculating
detection rates.  In Section III, we review the distribution functions
for the Evans models which we use to investigate the effects of
flattening and bulk rotation.  Results for the effects of flattening on
the local WIMP velocity distribution, density, and total and
differential detection rates are provided in Section IV.  
We also propose here that the measured differential
recoil-energy distribution (in case of detection) could be used
to constrain the bulk rotation and flattening of the halo.
In Section V
we investigate the effects of uncertainties in the halo radial profile
in spherical models on dark-matter detection rates.  In Section VI we
summarize and make some concluding remarks.  We also discuss how rates
for indirect detection of WIMPs will be affected in these
alternative halo models.

\section{Calculating Direct-Detection Rates}

One can write the differential rate for direct WIMP detection
\cite{jkg} as
\begin{equation}
     \frac{dR}{dQ}=\frac{\sigma_0 \rho_0}{2 m_\chi
     m_r^2}F^2(Q)\int_{v_{\text{min}}}^{\infty} {\frac{f_1(v)}{v}dv},
\end{equation}
where $\sigma_0$ is the cross-section (at zero momentum
transfer); $\rho_0$ is the local dark matter
density; $m_r$ is the reduced mass $m_N
m_\chi(m_N+m_\chi)^{-1}$, where $m_N$ is
the mass of a target nucleus and $m_\chi$ is the WIMP mass; $Q=|{\bf
q}|^2/2 m_N$ is the deposited energy, where ${\bf q}$ is the momentum transfer;
$F(Q)$ is a nuclear form factor; $f_1(v)$ is the distribution of WIMP
speeds relative to the detector (normalized to 1); and
$v_{\text{min}}=[(Q m_N)/(2 m_r^2)]^{1/2}.$
Defining the dimensionless quantity,
\begin{equation}T(Q)=\frac{\sqrt{\pi}}{2}v_0\int_{v_{\text{min}}}^{\infty}
{\frac{f_1(v)}{v}dv},
\end{equation}
and taking $F(Q)={\text{exp}}(-Q/2Q_0),$ the differential detection rate can be
written as
\begin{equation}
\frac{dR}{dQ}=\bigg(\frac{\rho_0 \sigma_0}{\sqrt{\pi}v_0 m_{\chi}
m_{\text{r}}^2}\bigg){\text{exp}}(-Q/Q_0) T(Q);
\end{equation}
i.e., the density times a velocity-dependent term.
The total event rate can be determined by integrating over all detectable
energies:
\begin{equation}
R=\int_{E_{\text T}}^{\infty}\frac{dR}{dQ}dQ,
\end{equation}
where ${E_{\text T}}$ is the threshold energy for the detector.

\section{HALO MODELS}

To study the effects of flattening and bulk rotation on
detection rates, we use Evans's family of analytic axisymmetric
distribution functions (DFs) \cite{evans},
\begin{equation}
F (E,L_z^2)=[A L_z^2+B] \text{exp}(4E/v_0^2)+C
\text{exp}(2E/v_0^2),
\label{evansdistribution}
\end{equation}
with
\begin{equation}
A\!=\!\bigg(\frac{2}{\pi}\bigg)^{5/2}\frac{(1-q^2)}{G q^2 v_0^3},\;
B\!=\!\bigg(\frac{2}{\pi^5}\bigg)^{1/2}\!\!\frac{R_c^2}{G q^2 v_0},\;
C\!=\!\frac{2q^2-1}{4\pi G q^2 v_0},
\end{equation}
where $E$ is the binding energy, $L_z$ is the azimuthal component of
angular momentum,
$v_0$ is the circular speed at large radii, $R_c$ is the core radius,
and $q$ is the flattening parameter, ranging from $q=1$ for a cored, spherical
halo to $q=1/\sqrt{2}\approx0.707$ for the most flattened non-negative
DF \cite{evans}.  These models elegantly
reproduce Binney's potential and corresponding density \cite{bt},
\begin{equation}
     \psi(R,z)=-\frac{1}{2} v_0^2
     \log{\bigg(R_c^2+R^2+\frac{z^2}{q^2}\bigg)},
\label{psiequation}
\end{equation}
\begin{equation}
     \rho(R,z)=\frac{v_0^2}{4 \pi G q^2}\frac{(2 q^2+1) R_c^2+R^2+(2-q^{-2})
     z^2}{(R_c^2+R^2+z^2 q^{-2})^2},
\label{rhoequation}
\end{equation}
where $R$ is the radial distance and $z$ is the vertical distance above the
disk.  These are suitable for describing the halo since they
produce rotation curves which rise linearly at small radii and
are flat at large radii.

\begin{figure}[htbp]
 \centerline{\psfig{file=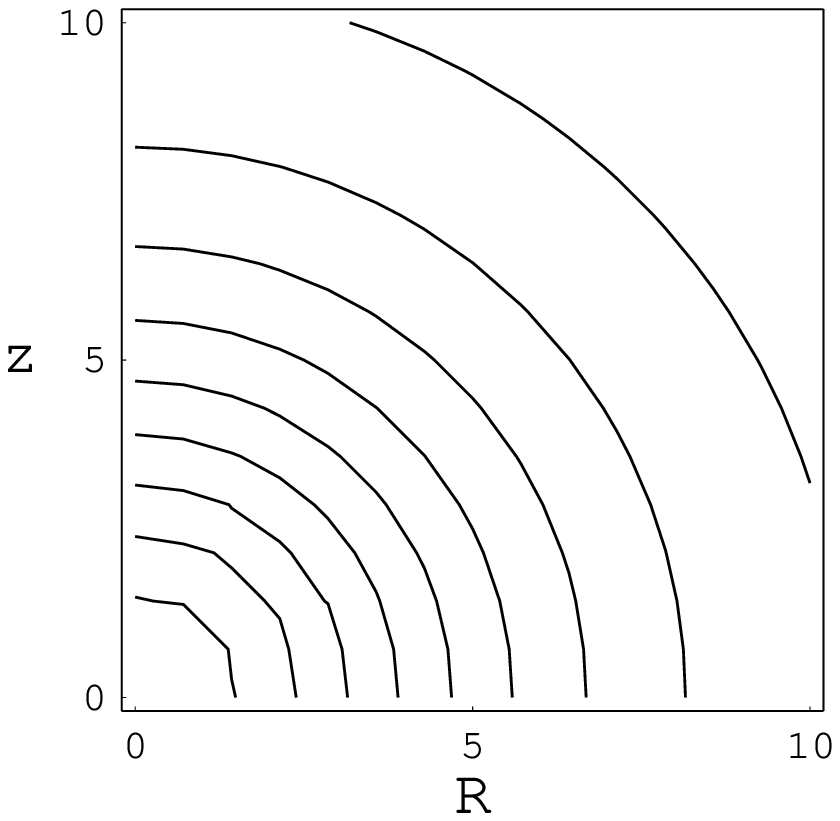,width=2.5in}}
 \centerline{\psfig{file=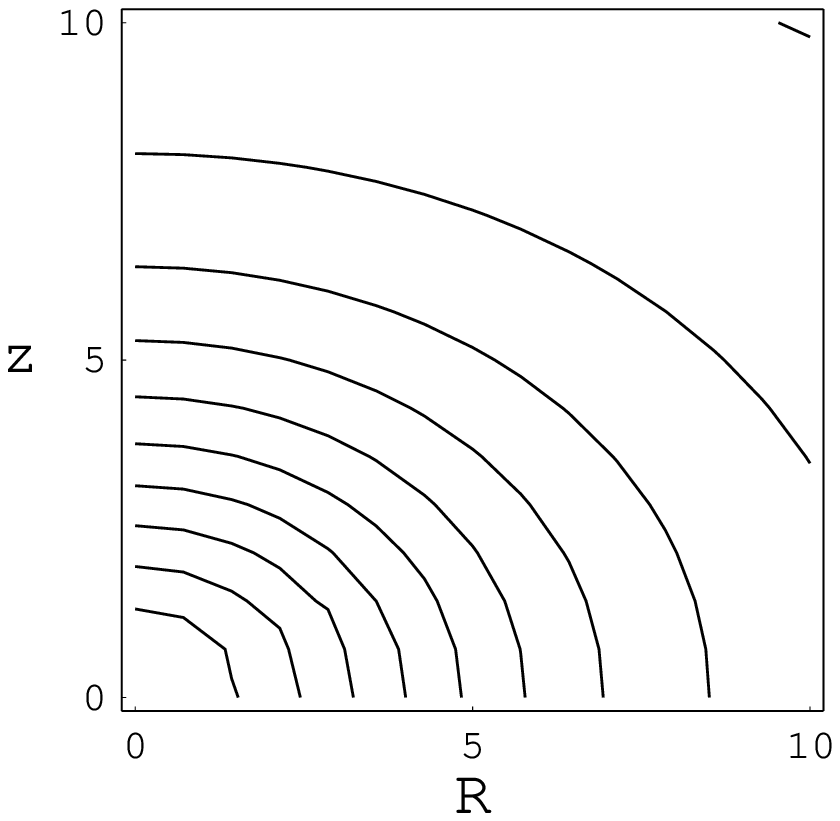,width=2.5in}}
 \centerline{\psfig{file=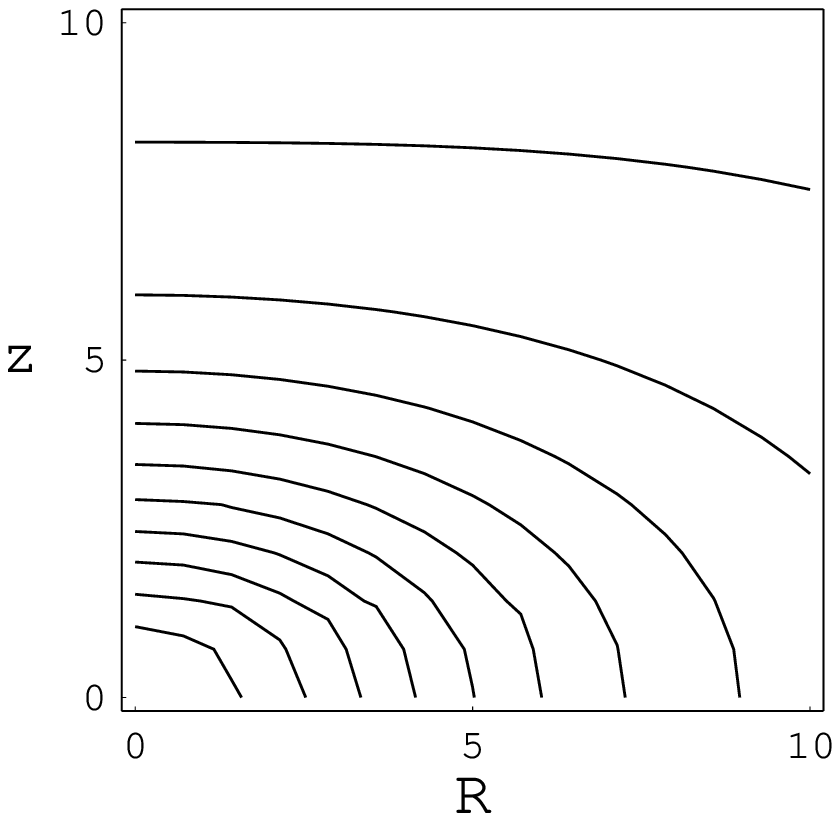,width=2.5in}}
 \bigskip
 \caption{Halo isodensity contours for the Evans models for
 $q=1,0.85,0.707$, where the $z$ and $R$ axes are in kpc.)} 
\label{equidensity}
\end{figure}

In this calculation, we take the dark-matter contribution to the
local circular velocity to be $v_c(R_0)=170$ km~sec$^{-1}$ (which
we get from a local rotation speed of 220 km~sec$^{-1}$ and a
disk contribution of 140 km~sec$^{-1}$),
$v_0=v_{\infty}=220$ km~sec$^{-1}$, a Galactocentric radius
$R_0=8.5$ kpc, and $z=0$ kpc.  The core radius $R_c$ is obtained by
noting that in the plane $z=0$,
\begin{equation}
     v_c^2=R\frac{d\psi}{dR}=\frac{v_0^2 R^2}{R_c^2+R^2}.
\end{equation}
Therefore, for all $q$, the core radius is
\begin{equation}
     R_c=R_0\bigg(\frac{v_\infty^2}{v_c(R_0)^2}-1\bigg)^{1/2}\approx 7\;
     {\text{kpc}}.
\end{equation}
[We have checked that our conclusions on the effects of
flattening are unchanged if we adopt other plausible values for
$v_\infty$, $v_c(R_0)$, and $R_0$.]  

The isopotential contours for these models are ellipsoidal with
(short-to-long) axis ratios $q$ [c.f., Eq. (\ref{psiequation})].
Fig. \ref{equidensity} shows isodensity contours for $q=1$, 0.85, and
$1/\sqrt{2}\approx 0.707$ (for $R_c=7$~kpc).  The isodensity
contours are not ellipsoidal.  For small radii, they are close
to spherical, and they become more flattened for larger radii.
Inspection of Fig. \ref{equidensity} shows that (for
$R_c=7$~kpc) the short-long axis ratio for the isodensity
contours is roughly 1:2 for $q\simeq0.707$ for radii
comparable to our Galactocentric radius.

The DFs above have no bulk rotation.  However, a family of
DFs with bulk rotations can be constructed by considering linear
combinations,
\begin{equation}
     G(E,L_z)=a F_+(E,L_z) + (1-a) F_-(E,L_z),
\end{equation}
of DFs
\begin{equation}
F_+(E,L_z^2)=
\cases{$$F(E,L_z^2)$$, & $v_\phi>0$;\cr
                    0, & $v_\phi<0$,\cr}
\end{equation}
\begin{equation}
F_-(E,L_z^2)=
\cases{             0, & $v_\phi>0$;\cr
       $$F(E,L_z^2)$$, & $v_\phi<0$,\cr}
\end{equation}
with only positive or negative azimuthal-velocity components
$v_\phi$.  These models have the same spatial distributions as
the nonrotating models $F(E,L_z)$.  The parameter $a$ ranges
{}from 1 (for maximal corotation) to 0.5 (the model with no net
rotation) to 0 (maximal counterrotation), and is related to the
dimensionless spin parameter $\lambda$ usually used to quantify
galactic angular momenta by $\lambda = 0.36|a-0.5|$.  

The DFs discussed so far specify the velocity distribution in
the Galactic rest frame.  However, the solar system moves with
respect to this frame with a velocity $v_s=220$ km~sec$^{-1}$.
Therefore, the DF $F_s(v_R,v_z,v_\phi)$ with respect to the Earth can be
obtained from the rest-frame DF $F$ by substituting $v_\phi
\rightarrow v_\phi+v_s$.

\subsection{No Net Rotation}

For these models, the distribution function is even in the variable $v_\phi$;
there are as many particles circling around clockwise as there are
counterclockwise.

Substituting the binding energy,
\begin{equation}
     E=-\frac{1}{2}v^2-\frac{1}{2} v_0^2 
     \log{\bigg(R_c^2+R^2+z^2/q^2\bigg)},
\end{equation}
into the distribution function $F(E,L_z^2)$ and transforming to the Sun's rest
frame yields
\begin{eqnarray}
F_s(E,L_z^2) &=& [A R^2 (v_\phi+v_s)^2+B]\nonumber\\
            &&
     \times \frac{\exp{\{-\frac{2}{v_0^2}
     (v_r^2+v_\theta^2+(v_\phi+v_s)^2)\}}}{(R_c^2+R^2+z^2/q^2)^2}\\
             &+C
     &\;\frac{\exp{\{-\frac{1}{v_0^2}
     (v_r^2+v_\theta^2+(v_\phi+v_s)^2)\}}}
     {(R_c^2+R^2+z^2/q^2)}.\nonumber
\end{eqnarray}

Since $v^2=v_R^2+v_z^2+v_\phi^2,$ one can simplify this to depend only on
$v$ and the angle $\alpha$ between the velocity and the
azimuthal direction.   Plugging in for the local
coordinates $(R,z)=(R_0,0)$, one obtains the more convenient form,
\begin{eqnarray}
     f(v,\alpha) &=& [A R_0^2 (v \cos\alpha+v_s)^2+B]\nonumber\\
      && \times\frac{\exp{\{-\frac{2}{v_0^2}(v^2+2v_s v \cos\alpha+v_s^2)\}}}{
      (R_c^2+R_0^2)^2}\\
      &+C&\;\frac{\exp{\{-\frac{1}{v_0^2}(v^2+2v_s v \cos\alpha+v_s^2)\}}}
      {(R_c^2+R_0^2)},\nonumber
\end{eqnarray}
where the $q$ dependence is still implicit in the coefficients.
Therefore, the local speed distribution function
needed for calculation of the dark-matter detection
rate is
\begin{equation}
     f_1(v)=\frac{\int_0^\pi\, {f(v,\alpha)\, v^2 \sin{\alpha}\,
     d\alpha}}{\int_0^\infty\, \int_0^\pi {f(v,\alpha)} \,v^2 \sin{\alpha}  \,d\alpha \,dv\,
     }.
\end{equation}

\begin{figure}[htbp]
\smallskip
   \centerline{\psfig{file=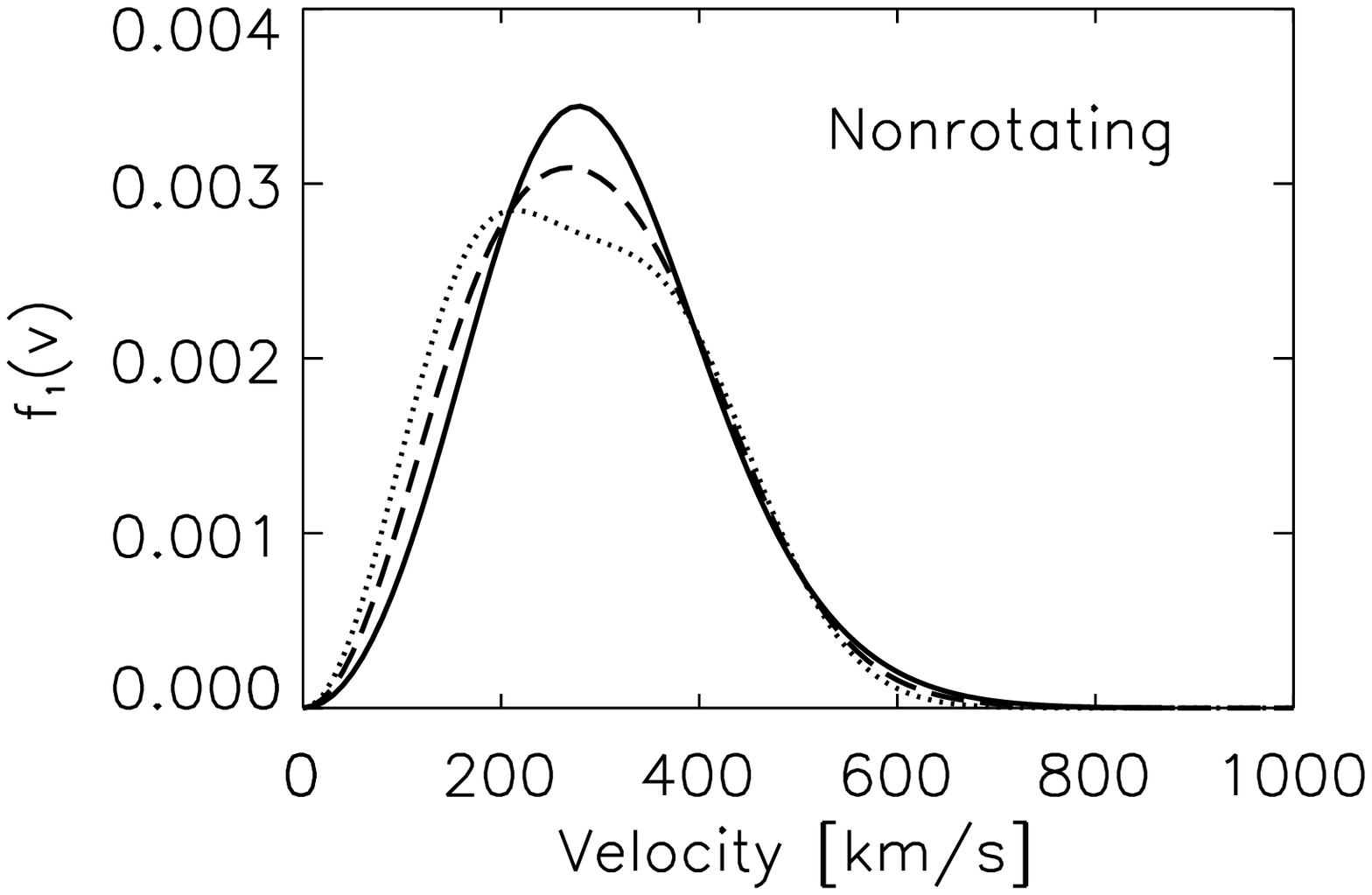,width=3.4in}}
   \centerline{\psfig{file=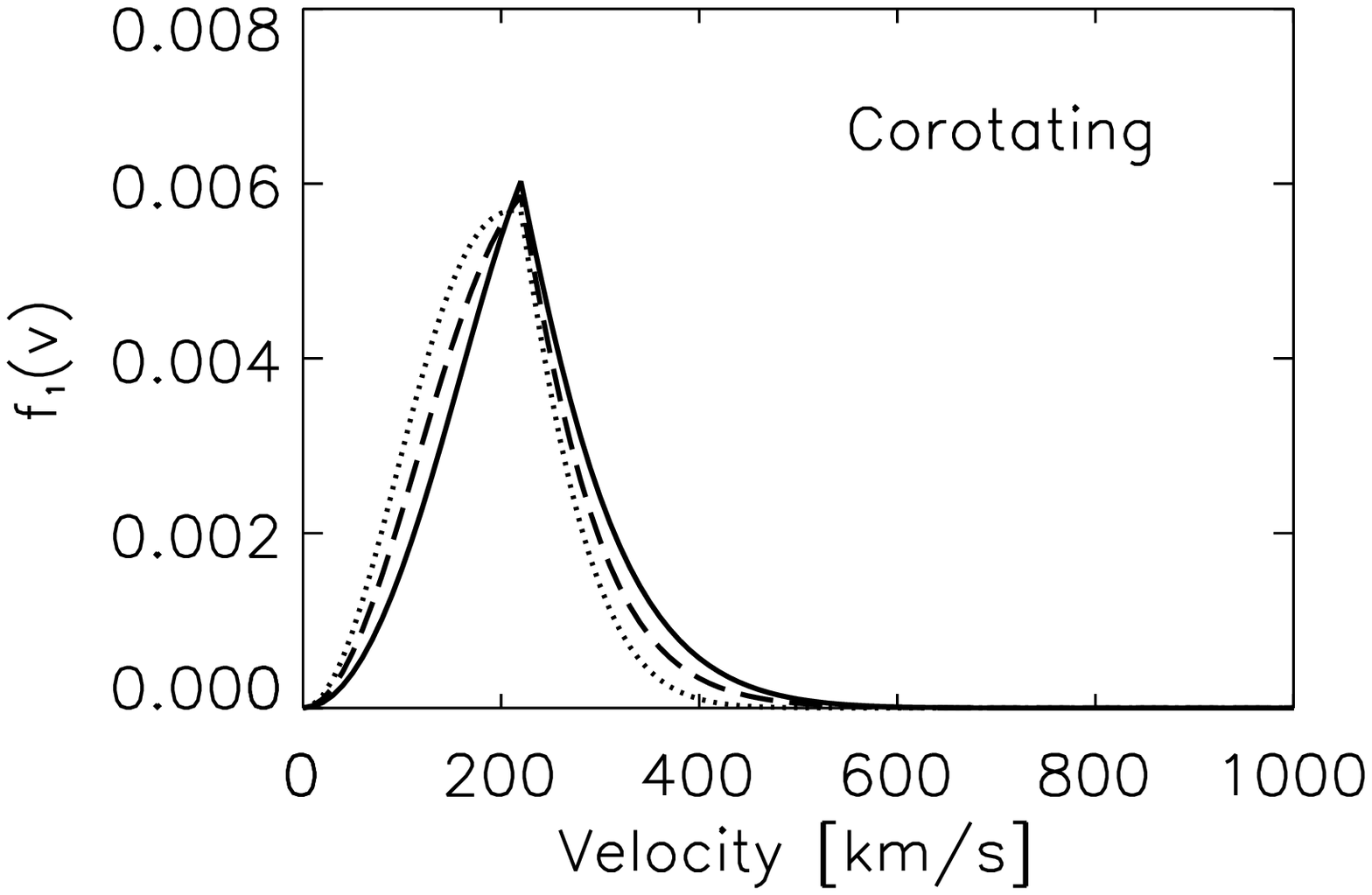,width=3.4in}}
   \centerline{\psfig{file=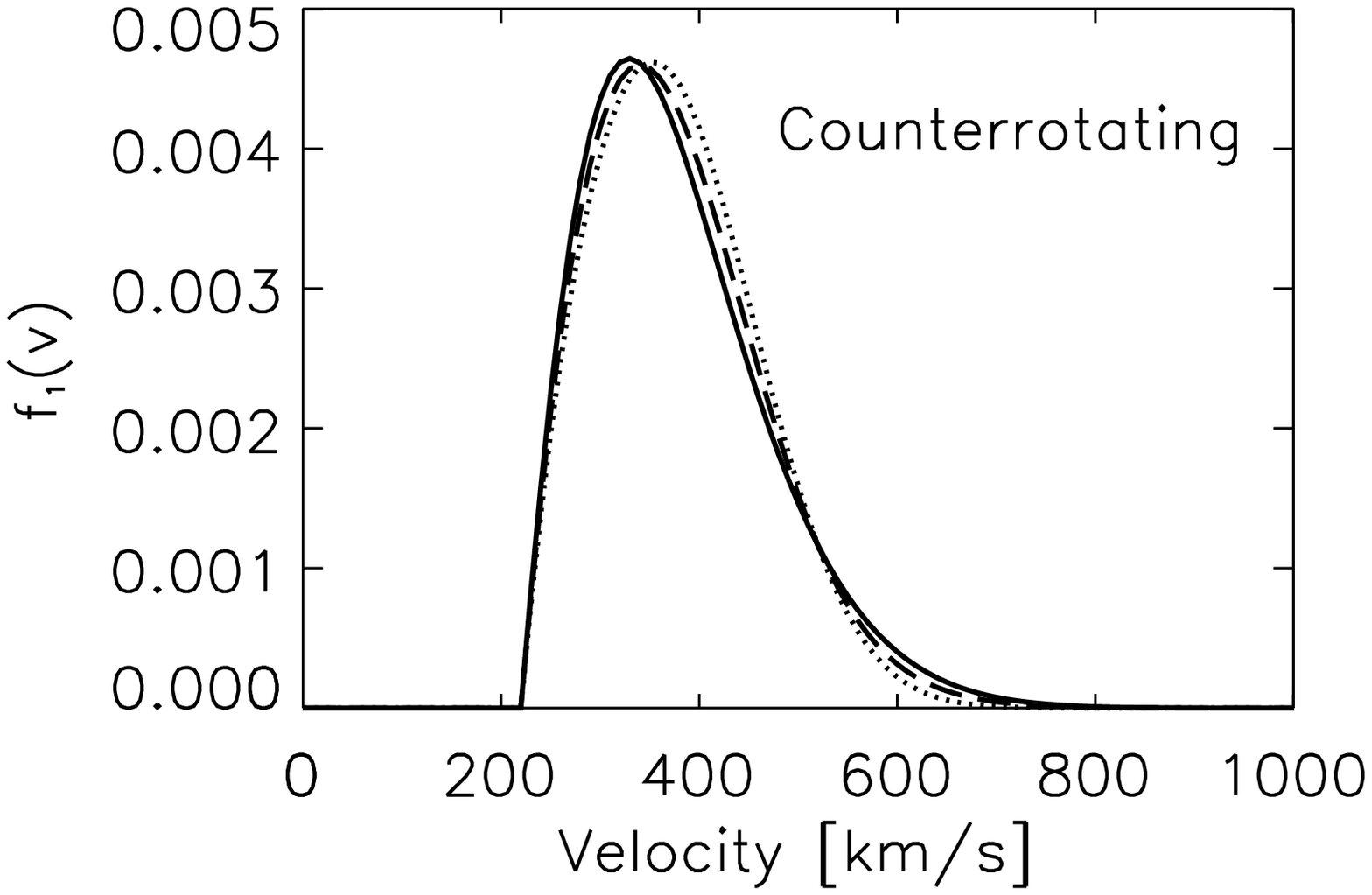,width=3.4in}}
 \medskip
 \caption{Local speed distributions $f_1(v)$ for nonrotating,
   maximally corotating, and maximally counterrotating
   models with $q=1$ (solid curves), $q=0.85$ (dashed
   curves), and $q=0.707$ (dotted curves) and with $v_s=220$ km/s.}
 \label{velocity}
 \end{figure}

The top panel in Fig. \ref{velocity} shows the speed
distributions $f_1(v)$ for the nonrotating halo for $q=1$, 0.85,
and 0.707.

\subsection{Maximally Corotating and Counterrotating}

The calculation of the speed distribution for a rotating halo
proceeds in the same fashion.  However, for the maximally
corotating case, the DF in the Sun's rest frame is
\begin{equation}
F_{+s}(E,L_z^2)=
\cases{$$F_s(E,L_z^2)$$, & $v_\phi>-v_s$;\cr
                    0, & $v_\phi<-v_s$,\cr}$$
\end{equation}
and for the maximally counterrotating model, the DF in the
Sun's rest frame is
\begin{equation}
F_{-s}(E,L_z^2)=
\cases{             0, & $v_\phi>-v_s$;\cr
       $$F_s(E,L_z^2)$$, & $v_\phi<-v_s$.\cr}$$
\end{equation}

The middle and bottom panels in Fig. \ref{velocity} show the
speed distributions
$f_1(v)$ for the maximally-corotating and counterrotating
models, respectively, again for $q=1$, 0.85, and 0.707.
Note that there are no particles with $v<v_s$ for the maximally
counterrotating model.  Also, the steep rise in $f_1(v)$ near
$v=0$ for the $q=0.707$ corotating model arises because there
are more particles in nearly circular orbits with velocities
$v_s$---nearer to $0$ in our frame---in this model than in the
$q=1$ model.  The maximally rotating models have
a spin parameter $\lambda=0.18$ which is significantly larger
than the spin parameters $\lambda\simeq0.05$ expected from
galaxy-formation models \cite{barnes}.  Therefore, realistic speed
distributions should lie somewhere between these two and closer
to that for the nonrotating model.

\section{TOTAL AND DIFFERENTIAL DIRECT-DETECTION RATES}

\begin{figure}[htbp]
 \centerline{\psfig{file=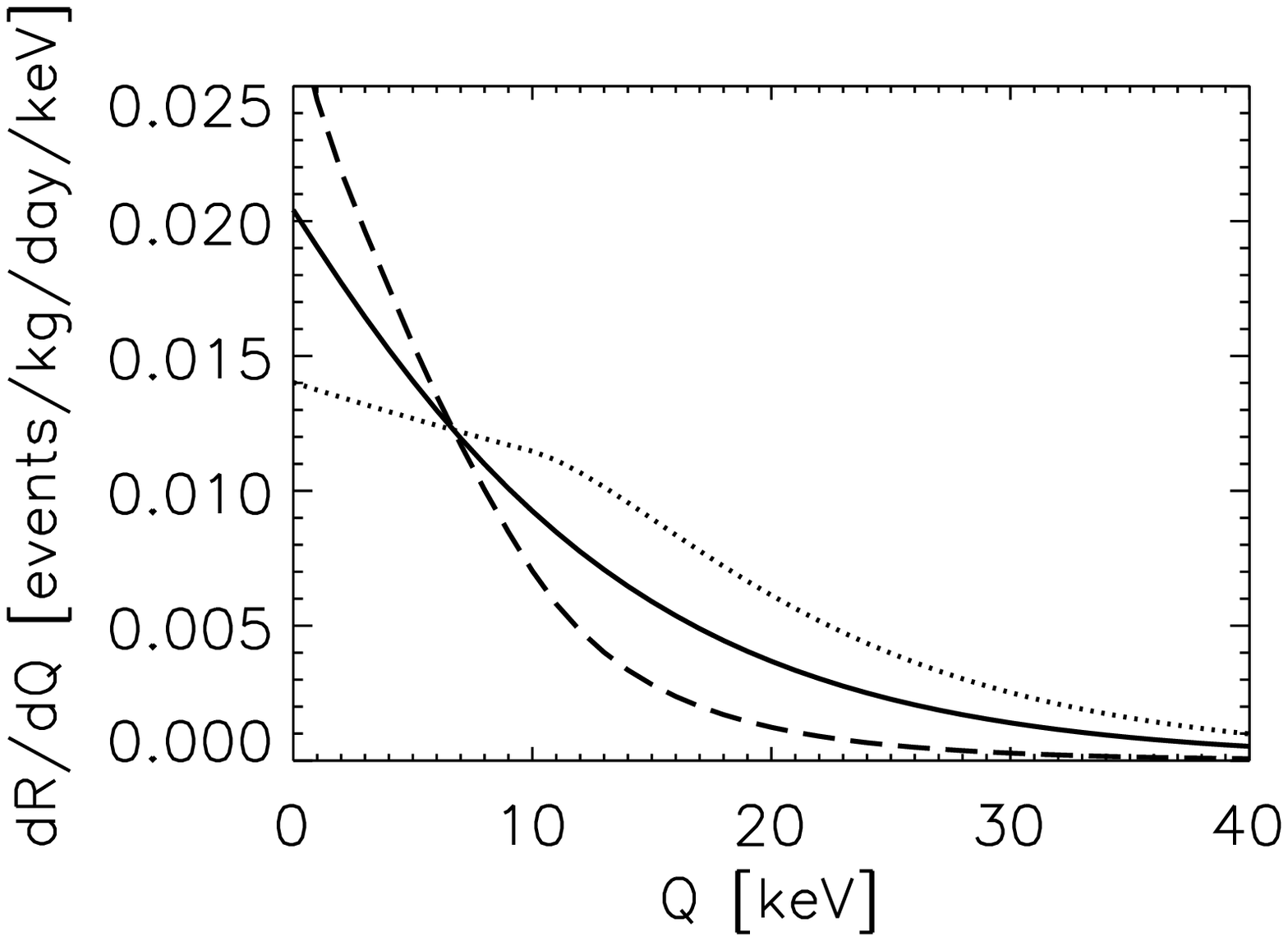,width=3.4in}}
 \centerline{\psfig{file=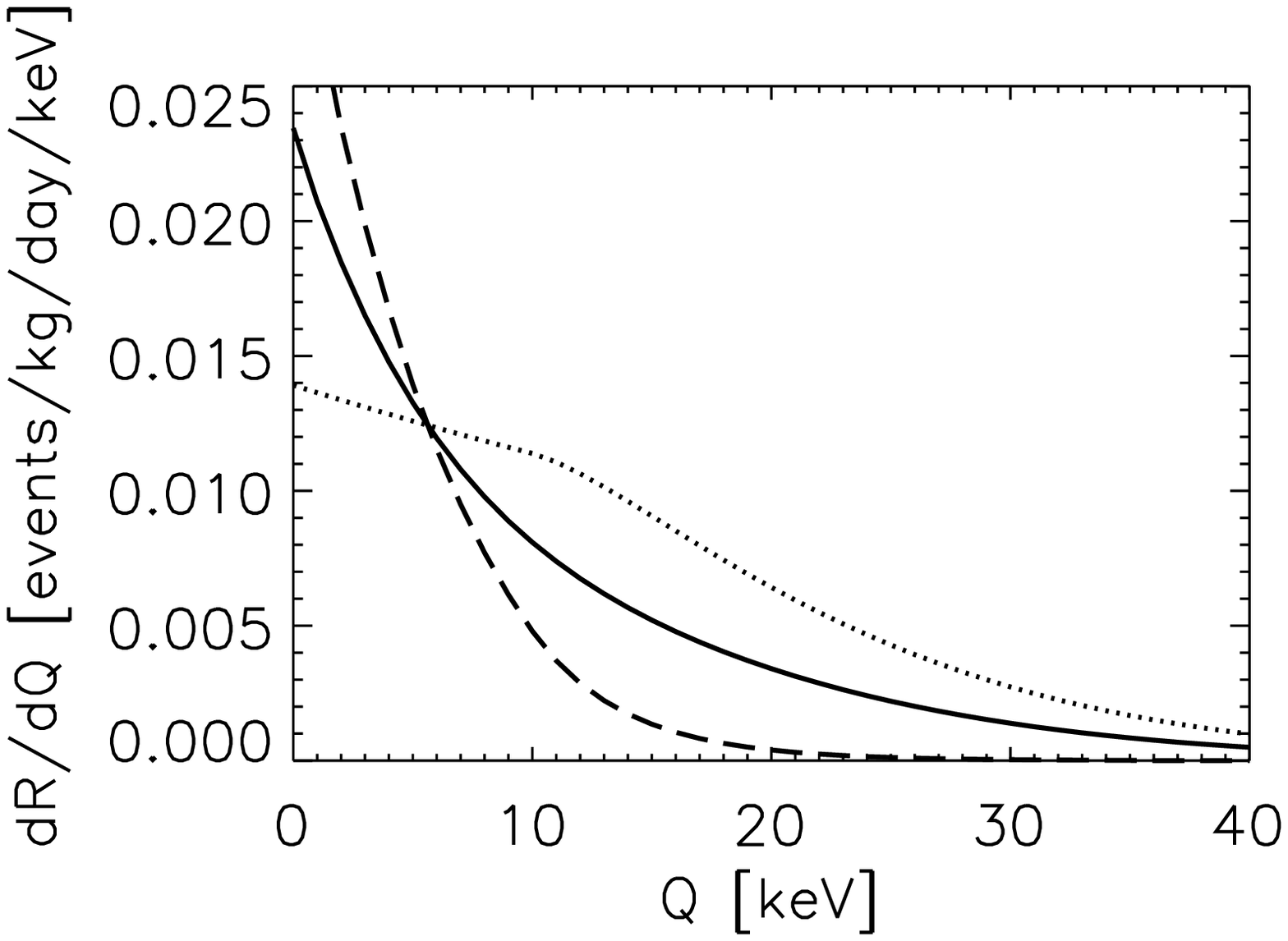,width=3.4in}}
 \bigskip
 \caption{Differential detection rates for Evans's models $q=1$ (top)
 and $q=0.707$ (bottom) with no net rotation (solid), maximal
 corotation (dashed), and maximal counterrotation (dotted).}
\label{dRdQ}
\end{figure}

Fig. \ref{dRdQ} shows the differential detection rates $dR/dQ$
for spherical and flattened nonrotating and maximally corotating
and counterrotating models.  It is seen that flattening has a
weak effect on the predicted differential-detection rate.  Bulk
rotation (especially counterrotation) has a somewhat stronger
effect on the differential rates.  Therefore, the shape of the
nuclear recoil spectrum could provide information on whether the
halo is rotating or not, and this could be useful for
constraining galaxy-formation models.

\begin{figure}[htbp]
 \centerline{\psfig{file=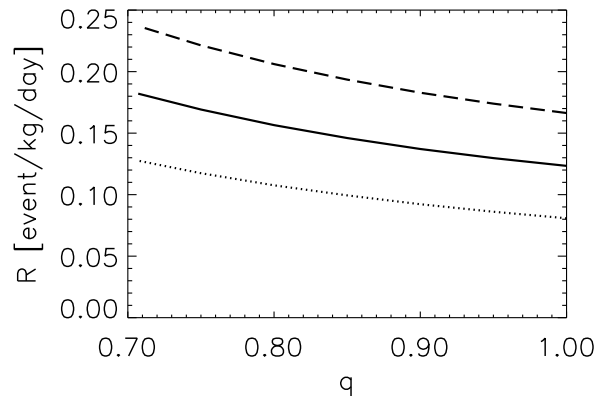,width=3.4in}}
 \medskip
 \caption{Total detection rate as a function of halo flattening $q$ for
 nonrotating (solid), maximally corotating (dashed), and maximally
 counterrotating (dotted) halos.}
\label{eventrate}
\end{figure}

Fig. \ref{eventrate} shows the total detection rate (assuming no
thresholds) for nonrotating and maximally corotating and
counterrotating models as a function of the flattening
parameter $q$.  The detection rate increases roughly as
$q^{-1}$ independent of the rotation.  The larger incident WIMP
velocities in counterrotating models leads to a stronger
form-factor suppression.  This is the leading factor in
accounting for the decrease in the event rate in
counterrotating models and {\it vice versa} for corotating
models.   Maximal rotation can change the event
rates by roughly 30\%.  However, the spin parameters expected on
theoretical grounds are generally smaller than a third of that
for our maximally rotating halos.  Therefore, the most plausible
values for the bulk rotation should yield detection rates within
10\% of those for the canonical nonrotating model.

\begin{figure}[htbp]
 \centerline{\psfig{file=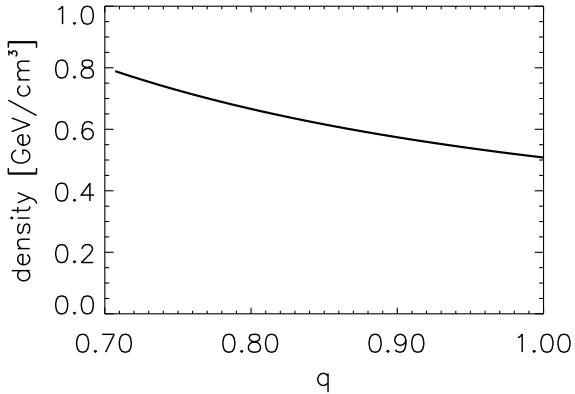,width=3.4in}}
 \medskip
 \caption{Local halo density as a function of the  flattening $q$.}
\label{density}
\end{figure}

\begin{figure}[htbp]
 \centerline{\psfig{file=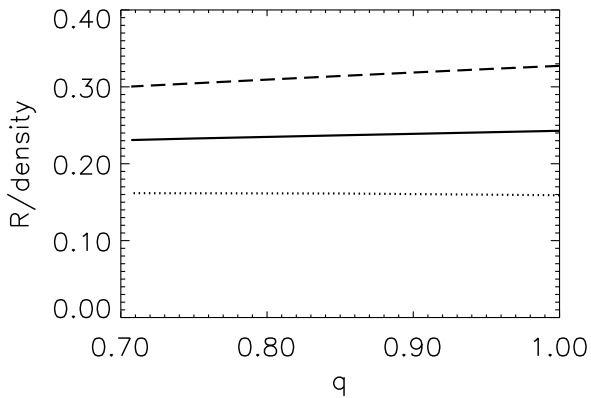,width=3.4in}}
 \medskip
 \caption{Velocity dependence (i.e., the detection rate scaled by the
 local halo density) of the total detection rate as a function of the
 flattening $q$ for nonrotating (solid), maximally corotating (dashed),
 and maximally counterrotating (dotted) halos.} 
\label{velcomponent}
\end{figure}

The $q$ dependence of the local halo density can be derived from
Eq. (\ref{rhoequation}), and Fig. \ref{density} shows that it scales
very nearly as $q^{-1}$.  Fig. \ref{velcomponent} shows the
detection rate scaled by the local halo density as a function of
$q$.  These two Figures illustrate that the change in
the velocity distribution from flattening has essentially no
effect on the dark-matter detection rate.  Heuristically,
halo particles move in the same gravitational potential as the
Sun, and the velocity dispersion of any species is fixed by the
potential.  Our calculations contradict the claims of Cowsik et
al. \cite{cowsik} and verify the arguments of
Refs. \cite{cowsikcritique}.

For these calculations, we have used a WIMP with only scalar
interactions of mass $m_\chi=100$ GeV and
$\sigma_0=4\times10^{-36}$ cm$^2$ and a Germanium target
nucleus.  We have checked that our conclusions do not change if
we use a different WIMP mass and/or target nucleus. We have also
checked that this conclusion is independent of the details of
the assumed rotation curve: The velocity dispersion is essentially
independent of the flattening in models where the halo contribution to
the local rotation curve is higher or lower than that which we have used
here, either because of different measured rotation speeds, or
because of a different disk/bulge contribution.

\section{RADIAL PROFILE}

Let us now consider the effect of possible variation in the
radial profile in spherical halo models.  Heuristically, galaxy
formation results in a cored isothermal halo through the process
of violent relaxation.  However, there will realistically be
some deviations from this simple physical picture for halo
formation.  For example, the collapse of baryonic matter in the
Galaxy might affect this process.  Empirically, the evidence for
flattened spiral-galaxy halos suggests some departure from the
simple picture.  Therefore, even if we consider only
spherical halo distributions, there is still some latitude in
our choice of the precise form for the radial profile of the
halo.

\begin{figure}[htbp]
 \centerline{\psfig{file=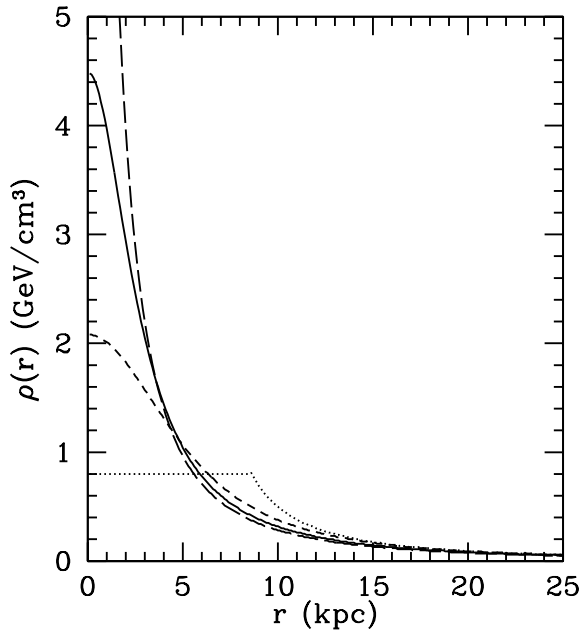,width=3.4in}}
 \bigskip
 \caption{The radial profile of the three spherical halo models
 discussed in the text.  The solid, short-dash, and long-dash
 curves are for the canonical isothermal
 [Eq. (5.1)], spherical Evans
 [Eq. (5.2)], and alternative isothermal
 [Eq. (5.3)] models, respectively.  The dotted curve is
 for the non-increasing radial profile that gives the largest local
 density.
 }
\label{rhocompare}
\end{figure}

\begin{figure}[htbp]
 \centerline{\psfig{file=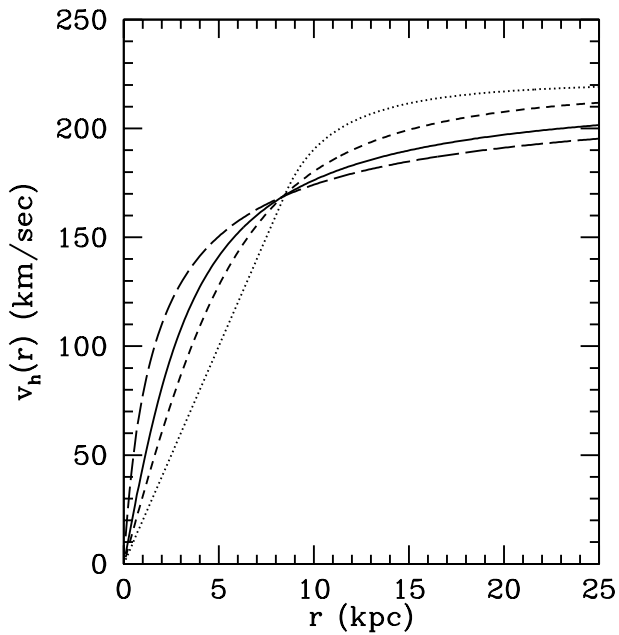,width=3.4in}}
 \bigskip
 \caption{Rotation curves for the spherical models shown in Fig.
 7.}
\label{vcompare}
\end{figure}

An empirically plausible radial profile for a spherical Galactic
halo is constrained by its contribution to the Galactic rotation
curve.  Therefore, it should approach a constant near the core
so it gives rise to a linearly rising rotation curve at small
radii, and it should fall as $r^{-2}$ at large radii to provide
a flat rotation curve.  The canonical profile
usually used for dark-matter calculations is the so-called
``isothermal'' sphere (actually, the radial profile of the true
cored isothermal sphere cannot be written analytically; see
Ref. \cite{bt}, p. 229),
\begin{equation}
     \rho(r) = {v_\infty^2 \over 4 \pi G r_c^2} {r_c^2 \over r_c^2 +
     r^2},
\label{isothermal}
\end{equation}
where $r_c$ is a core radius which is fit to the halo contribution
to the local rotation speed (and $r$ is now the spherical radial
coordinate, $r^2=R^2+z^2$).  Of course, the radial profile of
the spherical Evans model,
\begin{equation}
     \rho(r) = {v_\infty^2 \over 4\pi G} {r^2+3 r_c^2 \over (r^2
     + r_c^2)^2},
\label{evansmodel}
\end{equation}
also has the desired properties.  Yet another analytic form
which might be empirically acceptable is
\begin{equation}
     \rho(r) = {v_\infty^2 \over 4 \pi G r_c^2} {r_c^2 \over (r_c +
     r)^2}.
\label{third}
\end{equation}
Keep in mind that the core radius $r_c$ for each model must be
fit to the rotation curve, and $r_c$ for each model will be
different.  Suppose, as we did before, that
the local rotation speed is 220 km~sec$^{-1}$ and the disk
contribution is 137 km~sec$^{-1}$.  Then the local halo
contribution to the rotation curve is 170 km~sec$^{-1}$ which
leads to core radii $r_c=7$ kpc (as before) for the Evans model,
$r_c=2.8$ kpc for the canonical isothermal sphere, and
$r_c=0.9$ kpc for the alternative in Eq. (\ref{third}).  The
rotation curves and radial profiles for these three models are
plotted in Figs. \ref{rhocompare} and \ref{vcompare}
respectively.  The solid, short-dash, and long-dash curves are
for Eqs. (\ref{isothermal}), (\ref{evansmodel}), and
(\ref{third}), respectively.

In principle, the radial profile can be fixed by the halo
contribution to the rotation curve.  However, measurement of the
Galactic rotation curve is notoriously difficult, especially
near the interior.  Furthermore, the disk contribution to the
rotation curve must be known to infer the halo contribution, and
precise determination of the disk contribution is also
difficult.  Therefore, there will be significant uncertainties
in any reconstruction of the halo contribution to the rotation
curve from observational data.

Using our canonical values for the halo contributions to the
rotation speed, we find local halo densities of
0.43,\footnote{This corrects the value of 0.35 GeV~cm$^{-3}$
given in Section 2.4 of Ref. \cite{jkg}.} 0.51, and
0.38 GeV~cm$^{-3}$ for the isothermal, Evans, and alternative
models respectively.  Therefore, although the central density of
these three models may differ considerably
(c.f. Fig. \ref{rhocompare}), the requirements that each yield
the same halo contribution to the local rotation speed and the
same asymptotic rotation speed constrain the local halo density
in these models to 20\%.

One could contemplate a profile with a smaller local
density with a higher central density. However, a profile with a
central density much greater than that in our alternative
model will have a core density comparable to the Bulge density
(approximately 50 GeV~cm$^{-3}$ \cite{zhao}) and will therefore
contradict observed Bulge dynamics.  It is therefore unlikely
that the local halo density can be reduced while maintaining the
same halo contribution to the local and asymptotic rotation
speeds.  Contrariwise, one could consider
a model with a larger local density and smaller core radius
which still gives the same contribution to the local halo
speed.  Any physically reasonable radial profile should be
monotonically decreasing with radius.  The limiting case (a
density which is constant interior to our Galactocentric radius;
the dotted curve in Figs. \ref{rhocompare} and \ref{vcompare})
yields a density 1.4~GeV~cm$^{-3}~[v_c(r_0)/v_\infty]^2$, which
results in 0.8 GeV~cm$^{-3}$ for a local halo rotation speed of 170
km~sec$^{-1}$.  Therefore, a local halo density roughly twice
that obtained from the canonical model is conceivable (although
perhaps somewhat artificial as indicated in
Fig. \ref{rhocompare}), and a local halo density 
$O(10\%)$ smaller than the canonical value is also possible.

We have evaluated numerically the direct-detection rate using
the DF, Eq. (\ref{evansdistribution}), for the spherical Evans
model and the Maxwell-Boltzmann velocity distribution for the
isothermal sphere.  We find that the detection rate with the
spherical Evans model is roughly 15\% larger than that in the isothermal
model.  Therefore, the difference in detection rates can be
attributed primarily to the difference in the local halo density
and only secondarily to the differences in the velocity
distribution.  Once again---as in the case of flattening---we
find that different radial profiles lead to roughly the same
velocity dispersions as long as both profiles are fit to the
same halo rotation speed.

\section{CONCLUSION}

Predictions for WIMP detection rates are almost always carried out
assuming the dark-matter distribution to be an isothermal
sphere.  When fit to reasonable
values of the halo contribution to the local and asymptotic
rotation speeds, the canonical isothermal halo gives a local halo
density 0.25--0.5 GeV~cm$^{-3}$.  Its velocity distribution is
Maxwell-Boltzmann with a velocity distribution fixed by the
asymptotic rotation speed.  

However, virtually all the empirical constraints to the halo
come from its observationally inferred contribution to the
Galactic rotation curve.  These (still rather poorly determined)
data are supplemented by some qualitative theoretical notions
about the halo: i.e., that it should be more diffuse than the
disk and monotonically decreasing with Galactocentric radius.
Many halo distributions can satisfy these observational and
theoretical constraints and still produce the same local and
asymptotic rotation speeds.

In this paper, we have calculated WIMP direct-detection rates in
several plausible alternatives to the canonical model.
We find that if the halo is flattened with an isopotential axial
ratio $q$, the direct-detection rate will increase by roughly $q^{-1}$.
This increase is due primarily to the effect of flattening on
the local halo density, which also increases as $q^{-1}$.  We
have used a self-consistent distribution function for a
flattened halo to verify that the effects of flattening on the
velocity distribution have virtually {\it no} effect on the
detection rate.  Local stellar kinematics and the thickness of
gas layers suggest that halo isodensity contours may be
flattened by up to a factor of 2 \cite{rob} corresponding to
$q=0.707-1$ for the Evans models, which would suggest that
flattening might increase the local halo density, and therefore
direct-detection rates, by a factor of 1.4.  However, the
heuristic argument that flattening should affect the detection
rate primarily through its effect on the density, and only
secondarily through its effect on the velocity distribution
should also apply to a halo with ellipsoidal isodensity (rather
than isopotential) contours.  In such models, the local density
is increased by a factor near 2 for a flattening near 2
\cite{gyuk}.

There are no empirical constraints to the bulk rotation of the
halo.  A maximally corotating or counterrotating halo could increase or
decrease the detection rate by 40\%.  However, galaxy-formation
scenarios generally predict bulk rotations no more than 0.3 of
maximal.  Therefore, we do not expect bulk rotation to change
the predicted event rates by more than 10\%.  Although simple
galaxy-formation models suggest that a halo would corotate if
it rotated at all, the existence of counterrotating disks
\cite{counter} suggests that a counterrotating halo might also be
plausible.

We found that in spherical models, the local density could be
increased by up to a factor of two and decreased slightly with
different radial profiles that still give the same local and
asymptotic halo rotation speeds.  In this work, we focussed on
halos with axial symmetry, but it is possible that the 
halo may deviate somewhat from axial symmetry.  However,
detection rates in reasonable triaxial models also generally
fall within a factor of two of the canonical detection rates
\cite{spergel}.

We restricted our analysis to direct detection.  However,
similar conclusions should apply to
rates for indirect detection of WIMPs via observation of
energetic neutrinos from WIMP annihilation in the Sun and/or
Earth.  Like direct-detection rates, these rates are controlled
primarily by the local halo density.  Since the velocity
dispersion is fixed to a large extent by the local and
asymptotic rotation speeds, indirect-detection rates should not
be affected by their dependence on the velocity distribution.

On the other hand, plausible deviations from the canonical
isothermal sphere can lead to dramatically different fluxes of
anomalous cosmic-ray antiprotons, positrons, and gamma rays from
WIMP annihilation in the halo.  These fluxes are determined by
an integral of the {\it square} of the density over the entire
halo.  Although the local halo density does not differ too much
in alternative models, the core density can differ
dramatically.  In particular, Fig. \ref{rhocompare} shows that
the central density can be increased perhaps by an order of
magnitude over that in the canonical model.  If so, then the
flux of gamma rays from WIMP annihilation in the Galactic center
would be increased by a factor of 100 over the fluxes predicted
in canonical models.

There is also the possibility that if WIMPs are detected, the
nuclear recoil spectrum might tell us about the structure of the
halo.  Fig. \ref{dRdQ} shows how the recoil spectrum could be
used to constrain the rotation of the halo.  We have also
investigated the magnitude of annual modulations in the event
rate due to the Earth's orbital motion around the Sun.  We found
a maximally corotating halo could increase the annual modulation
by a factor of 2, implying an increase in modulation amplitude of 
$O(30\%)$ for models with more realistic corotation.

\acknowledgments

We thank M. Weil for useful discussions.  M.K. was supported by
the D.O.E grant number DEFG02-92-ER 40699, NASA NAG5-3091, and
the Alfred P. Sloan Foundation.   A.K. was supported by the
Columbia Rabi Scholars Program which is funded in full by the
Kann Rasmussen Foundation.

\end{document}